\documentclass{ws-ijmpa}
\usepackage[a4paper]{geometry}
\usepackage[super]{cite}
\usepackage{xcolor}
\usepackage[verbose,hypertexnames=false]{hyperref}
\usepackage{url}
\usepackage{bm}
\hypersetup{colorlinks=false,allbordercolors=blue,pdfborderstyle={/S/U/W 1}}

\begin{document}

\markboth{Daniel Arturo López Aguilar}{CP violation in two meson tau decays}

%
\catchline{}{}{}{}{}
%

\title{CP violation in two meson tau decays}

\author{Daniel Arturo López Aguilar}

\address{Departamento de Física, Centro de Investigación y de Estudios Avanzados del Instituto
Politécnico Nacional Apartado Postal 14-740, 07360 Ciudad de México, México \\
daniel.lopez.aguilar@cinvestav.mx}

\maketitle

\begin{abstract}
CP violation in $\tau\to K_S \pi\nu_\tau$ decays has attracted a lot of attention recently, due to the BaBar anomaly in the corresponding rate asymmetry. Within an effective field theory formalism, only extreme fine-tuning would allow to understand such measurement, which is currently being scrutinized at Belle(-II), as will be in the future super-charm-tau factory. Here we summarize the results of applying the same formalism to the other two-meson tau decay channels, which can help solve this conundrum. Our main conclusion is that current and future experiments would be sensitive to the maximum allowed CP rate asymmetry in the related $K^\pm K_S$ modes with a measurement having 5$\%$ precision, that will either support or cast further doubts on the BaBar anomaly.\end{abstract}


\ccode{PACS numbers: 11.30.Er, 12.39.Fe, 12.60.-i, 13.35.Dx}

\section{Introduction}
\hspace{0.5cm}As put forward by Zakharov long ago~\cite{Sakharov:1967dj}, violating the invariance under the combined symmetry operations of charge conjugation (C) and parity (P), CP, is a required ingredient~\footnote{A Lorentz invariant local quantum field theory with Hermitian Hamiltonian is assumed.}  to understand the baryon asymmetry of the Universe. Baryon number violation (perhaps leptonically-generated~\cite{Fukugita:1986hr}), C-violation, and non-thermal equilibrium interactions are also needed, although met in the Standard Theory (SM)~\cite{Glashow:1961tr, Weinberg:1967tq, Salam:1968rm}. This extremely successful, predictive and accurate framework is, however, insufficient to provide the current matter-antimatter asymmetry~\cite{ParticleDataGroup:2024cfk}. This is one of the main puzzles that must find a solution within quantum field theory and particle physics. As such, searches for additional sources of CP violation (CPV), and their understanding are a top priority nowadays.

Semileptonic tau decays are a clean environment for learning about non-perturbative QCD, and also quite a useful instrument in new physics searches~\cite{Pich:2013lsa, Davier:2005xq}. The interest of CPV studies in these decays was emphasized at BaBar and Belle(-II)~\cite{Tsai:1996ps, Kuhn:1996dv, Bigi:2012km, Bigi:2012kz, Kiers:2012fy} and recently for the super-tau-charm factory~\cite{Cheng:2025kpp}. CPV in hadron tau decays received much more attention since the anomaly reported by BaBar \cite{BaBar:2011pij} on the normalized asymmetry between the decay rates of a $\tau^+$ and a $\tau^-$ into final states with a prong pion and a $K_S$ (the measurement was inclusive in neutral pions). This CPV observable is predicted to be $(0.36\pm0.01)\%$ within the SM, using the extremely well-known neutral kaon mixing and accounting for the precise BaBar experimental setting~\cite{Bigi:2005ts,Calderon:2007rg,Grossman:2011zk}. Surprisingly, the measurement found the opposite sign (right magnitude), $-(0.36 \pm 0.23 \pm 0.11)\%$ (where errors are of statistical and systematic origin, respectively), at $2.8$ standard deviations from the SM prediction. Refs. \citen{Devi:2013gya,Dhargyal:2016kwp,Dighe:2019odu} claimed that new physics coupling to an antisymmetric tensor current could explain this anomaly~\footnote{We note that the K\"uhn-Mirkes formalism~\cite{Kuhn:1992nz} needs to be generalized to study these effects in $\tau\to K_S \pi \pi^0\nu_\tau$ decays~\cite{Arteaga:2022xxy,Aguilar:2025enl}.}. However, the equality of the tensor and vector form factors phases in the elastic region challenged this solution, barring extreme fine-tuning \cite{Cirigliano:2017tqn} (see also Ref. \citen{Rendon:2019awg,Chen:2019vbr,Chen:2020uxi,Sang:2020ksa,Chen:2021udz})~\footnote{The binned Belle analysis~\cite{Belle:2011sna} of these decays was compatible with CP due to the larger uncertainties, preventing a check of the BaBar result.}.

Here we summarize the main findings of our paper~\cite{Aguilar:2024ybr}, extending previous studies to the other two-meson tau decays of interest ($\pi\pi$, $K\bar{K}$ and $K\pi^0$ channels~\footnote{The $\pi\eta^{(\prime)}$ channels have not been measured yet, and the corresponding hadron uncertainty is big~\cite{Descotes-Genon:2014tla, Escribano:2016ntp}, so any present prediction is subject to big errors. CP conserving new physics signals in these decays were studied in \cite{Garces:2017jpz}. We will finally comment briefly also on the $K(\eta/\eta')$ channels, again subject to large uncertainties.}).
As heavy new physics contributions to the CPV $\tau\to K_S\pi\nu_\tau$ observables is strongly restricted, model-independently, using effective-field theory (EFT) \cite{Buchmuller:1985jz, Grzadkowski:2010es} to relate different energy scales complying with symmetry; it will happen analogously in the other two-meson tau decays. We compute the maximum amount of CPV possibly due to heavy new physics in them. Thus, any observation beyond our predicted bounds~\footnote{Within the SM CPV is negligible in modes without a neutral Kaon~\cite{Delepine:2005tw}.} will either be caused by light new physics or point to underestimated uncertainties. Testing these observables is pressing at Belle-II \cite{Belle-II:2018jsg} and future super-tau-charm factories~\cite{Achasov:2023gey, Achasov:2024eua,Cheng:2025kpp}.
\section{Low-energy Effective Field Theory Computation}\label{sec:LEFTCPVObservables}
The most general low-energy effective Lagrangian for the $\tau^-\to\bar{u}D\nu_\tau\, (D=d,s)+c.c.$ decays is given, at operator dimension six, by ~\footnote{We use $\sigma^{\mu\nu}=\frac{i}{2}[\gamma^\mu,\gamma^\nu]$ and $\epsilon^{0123}=+1$. The global phase is unphysical.}\cite{Cirigliano:2009wk,Garces:2017jpz,Cirigliano:2017tqn}
\begin{eqnarray}\label{LEFTLagrangian}
\mathcal{L}_{
EFT}=-\frac{G_F^0V_{uD}}{\sqrt{2}}\left(1+\epsilon_L^{D}+\epsilon_R^{D}\right)\Bigg\lbrace\bar{\tau}\gamma_\mu(1-\gamma_5)\nu_\tau\cdot\bar{u}\left[\gamma^\mu-\left(1-2\hat{\epsilon}_R^{D}\right)\gamma^\mu\gamma_5\right]D\nonumber\\
+\bar{\tau}(1-\gamma_5)\nu_\tau\cdot\bar{u}\left[\hat{\epsilon}_S^{D}-\hat{\epsilon}_P^{D}\gamma_5\right]D+2\hat{\epsilon}_T^{D}\bar{\tau}\sigma_{\mu\nu}(1-\gamma_5)\nu_\tau\cdot\bar{u}\sigma^{\mu\nu}D\Bigg\rbrace+\mathrm{h.c.}
\end{eqnarray}

From it, the decay amplitude for the $\tau^-\to K^- \pi^0 \nu_\tau$ decays reads
\begin{equation}\label{matrixelement}
\mathcal{M}(\tau^-\to K^- \pi^0 \nu_\tau)=\frac{G_FV_{us}}{2}\left[L_\mu H^\mu+\hat{\epsilon}_S^*LH+2\hat{\epsilon}_T^*L_{\mu\nu}H^{\mu\nu}\right]\,,
\end{equation}
where the lepton and hadron currents are
\begin{eqnarray}
   L&=&\bar{u}(p_{\nu_\tau})(1+\gamma_5)u(p_\tau)\,,\\
   L_\mu&=&\bar{u}(p_{\nu_\tau})\gamma_\mu(1-\gamma_5)u(p_\tau)\,,\\
   L_{\mu\nu}&=&\bar{u}(p_{\nu_\tau})\sigma_{\mu\nu}(1+\gamma_5)u(p_\tau)\,,
\end{eqnarray}
and
\begin{eqnarray}
H&=&\left\langle\pi^0(p_\pi)K^-(p_K)|\bar{s}d|0\right\rangle=\frac{\Delta_{K\pi}}{m_s-m_u}F_0(s)\,,\\
H^\mu&=&\left\langle\pi^0(p_\pi)K^-(p_K)|\bar{s}\gamma^\mu d|0\right\rangle=\left[(p_\pi-p_K)^\mu+\frac{\Delta_{K\pi}}{s}q^\mu\right]F_+(s)-\frac{\Delta_{K\pi}}{s}q^\mu F_0(s)\,,\;\;\;\;\;\;\\
H^{\mu\nu}&=&\left\langle\pi^0(p_\pi)K^-(p_K)|\bar{s}\sigma^{\mu\nu} d|0\right\rangle=iF_T(s)(p_\pi^\mu p_K^\nu-p_K^\mu p_\pi^\nu)\,,
\end{eqnarray}
in which $q^\mu=(p_\pi+p_K)^\mu$, $s=q^2$, $\Delta_{K\pi}=m_K^2-m_\pi^2$, and $F_{0,+,T}(s)$ are, respectively, the scalar, vector and tensor form factors, encapsulating the hadronization of the corresponding quark currents between the vacuum and the final-state mesons, in presence of strong interactions.
Our results for the selected set of observables verify those in Ref. \citen{Chen:2021udz}, namely:
\begin{eqnarray}
 &&\frac{\mathrm{d}^2\Gamma(\tau^-\to K^-\pi^0\nu_\tau)}{\mathrm{d}s\,\mathrm{d}\cos\alpha}=\frac{G_F^2|V_{us}|^2M_\tau^3S_{EW}}{1024\pi^3s^3}\left(1-\frac{s}{M_\tau^2}\right)^2\lambda^{1/2}(s,m_K^2,m_\pi^2)\nonumber\\
 &&\times\Bigg\lbrace\lambda(s,m_K^2,m_\pi^2)\left[\frac{s}{M_\tau^2}+\left(1-\frac{s}{M_\tau^2}\right)\cos^2\alpha\right]|F_+(s)|^2+\Delta_{K\pi}^2\Bigg|1+\frac{\hat{\epsilon}_S s}{M_\tau(m_s-m_u)}\Bigg|^2|F_0(s)|^2\nonumber\\
&&+4\lambda(s,m_K^2,m_\pi^2)\left[s|\hat{\epsilon}_T|^2\left(1-\left(1-\frac{s}{M_\tau^2}\right)\cos^2\alpha\right)|F_T(s)|^2-\frac{s}{M_\tau^2}\Re e[\hat{\epsilon}_TF_+(s)F^*_T(s)]\right]\nonumber\\
&&-2\Delta_{K\pi} \lambda^{1/2}(s,m_K^2,m_\pi^2)\Re e\left[\left(1+\frac{\hat{\epsilon_S s}}{M_\tau(m_s-m_u)}\right)F_+(s)F_0^*(s)\right]\cos\alpha\nonumber\\
&&+\frac{4s}{M_\tau}\Delta_{K\pi}\lambda^{1/2}(s,m_K^2,m_\pi^2)\Re e\left[\hat{\epsilon}^*_T\left(1+\frac{\hat{\epsilon}_S s}{M_\tau(m_s-m_u)}\right)F_T(s)F_0^*(s)\right]\cos\alpha\Bigg\rbrace\,, 
\end{eqnarray}
where $\lambda(a,b,c)=a^2+b^2+c^2-2(ab+ac+bc)$, and $S_{EW}\sim1.02$~\cite{Marciano:1988vm} encodes the universal short-distance electroweak radiative corrections~\footnote{We are not photon-inclusive, so (process-dependent) long-distance electromagnetic corrections~\cite{Cirigliano:2001er, Cirigliano:2002pv, Miranda:2020wdg,Cirigliano:2026ios} are not applied.}). Integrating the angular dependence (see our conventions in Ref. \citen{Kuhn:1992nz}), we get
\begin{eqnarray}
&&\frac{\mathrm{d}\Gamma(\tau^-\to K^-\pi^0\nu_\tau)}{\mathrm{d}s}=\frac{G_F^2|V_{us}|^2M_\tau^3S_{EW}}{1024\pi^3s^3}\left(1-\frac{s}{M_\tau^2}\right)^2\lambda^{1/2}(s,m_K^2,m_\pi^2)\nonumber\\
&&\times\Bigg\lbrace \frac{2}{3}\lambda(s,m_K^2,m_\pi^2)\left(1+\frac{2s}{M_\tau^2}\right)|F_+(s)|^2+2\Delta_{K\pi}^2\Bigg|1+\frac{\hat{\epsilon}_S s}{M_\tau(m_s-m_u)}\Bigg|^2|F_0(s)|^2\nonumber\\
&&+\frac{8}{3}\lambda(s,m_K^2,m_\pi^2)\left(s|\hat{\epsilon}_T|^2\left(2+\frac{s}{M_\tau^2}\right)|F_T(s)|^2-\frac{3s}{M_\tau}\Re e[\hat{\epsilon}_TF_+(s)F^*_T(s)]\right)\Bigg\rbrace\,.\;\;\;\;\;\;\;\;\;
\end{eqnarray}

Integrating over $s$, we obtain the partial decay width for any given channel.

Of main interest is the rate CP asymmetry:
\begin{eqnarray}\label{eq_ACPrateNP}
A^{\mathrm{rate}}_{CP}(\tau^\pm\to K^\pm\pi^0\nu_\tau)&=&\frac{\Gamma(\tau^+\to K^+\pi^0\bar{\nu}_\tau)-\Gamma(\tau^-\to K^-\pi^0\nu_\tau)}{\Gamma(\tau^+\to K^+\pi^0\bar{\nu}_\tau)+\Gamma(\tau^-\to K^-\pi^0\nu_\tau)}\nonumber\\
&=&\frac{\Im m[\hat{\epsilon}_T]G_F^2|V_{us}|^2S_{EW}}{128\pi^3M_\tau^2\Gamma(\tau\to K\pi^0\nu_\tau)}\int_{(m_K+m_\pi)^2}^{M_\tau^2}\mathrm{d}s\left(1-\frac{M_\tau^2}{s}\right)^2\lambda^{3/2}(s,m_K^2,m_\pi^2)\nonumber\\
&& \times |F_T(s)||F_+(s)|\sin[\delta_T(s)-\delta_+(s)]\,,
\end{eqnarray}
where $\delta_{+,T}(s)$ denote the phases of the vector and tensor form factors. The last lines of eq.~(\ref{eq_ACPrateNP}) are the new physics contribution to $A_{CP}^{rate}$. In the $\tau^\pm\to K^\pm K_S\nu_\tau$ case, this observable is dominated by the SM contribution, according to~\cite{Devi:2013gya}
\begin{equation}\label{eq_FullACPrate}
A_{CP}^{rate}\,=\,\frac{A_{CP}^{rate}|_{SM}+A_{CP}^{rate}|_{NP}}{1+A_{CP}^{rate}|_{SM}\times A_{CP}^{rate}|_{NP}}\,.
\end{equation}
Another useful observable is the forward-backward asymmetry,
\begin{equation}
A_{FB}^{\tau^-\to K^-\pi^0\nu_\tau}(s)=\frac{\int_0^1\frac{\mathrm{d}^2\Gamma(\tau^-\to K^-\pi^0\nu_\tau)}{\mathrm{d}s\,\mathrm{d}\cos\alpha}\mathrm{d}\cos\alpha-\int_{-1}^0\frac{\mathrm{d}^2\Gamma(\tau^-\to K^-\pi^0\nu_\tau)}{\mathrm{d}s\,\mathrm{d}\cos\alpha}\mathrm{d}\cos\alpha}{\int_0^1\frac{\mathrm{d}^2\Gamma(\tau^-\to K^-\pi^0\nu_\tau)}{\mathrm{d}s\,\mathrm{d}\cos\alpha}\mathrm{d}\cos\alpha+\int_{-1}^0\frac{\mathrm{d}^2\Gamma(\tau^-\to K^-\pi^0\nu_\tau)}{\mathrm{d}s\,\mathrm{d}\cos\alpha}\mathrm{d}\cos\alpha}\,,
\end{equation}
verifying $A_{FB}(s)=3/2<\cos\alpha>(s)$, where
\begin{equation}
<\cos\alpha>(s)=\frac{\int_{-1}^1\cos\alpha\left(\frac{\mathrm{d}^2\Gamma(\tau^-\to K^-\pi^0\nu_\tau)}{\mathrm{d}s\,\mathrm{d}\cos\alpha}\mathrm{d}\cos\alpha\right)}{\int_{-1}^1\left(\frac{\mathrm{d}^2\Gamma(\tau^-\to K^-\pi^0\nu_\tau)}{\mathrm{d}s\,\mathrm{d}\cos\alpha}\mathrm{d}\cos\alpha\right)}=\frac{N(s)}{D(s)},
\end{equation}
with
\begin{eqnarray}\label{eq_N(s)}
    N(s)=&-&\frac{4}{3}\Delta_{K\pi}\lambda^{1/2}(s,m_K^2,m_\pi^2)\Re e\left[\left(1+\frac{\hat{\epsilon}_S s}{M_\tau(m_s-m_u)}\right)F_+(s)F_0^*(s)\right]\\
    &+&\frac{8s}{3M_\tau}\Delta_{K\pi}\lambda^{1/2}(s,m_K^2,m_\pi^2)\Re e\left[\hat{\epsilon}^*_T\left(1+\frac{\hat{\epsilon}_S s}{M_\tau(m_s-m_u)}\right)F_T(s)F_0^*(s)\right]\,,\;\;\;\;\;\;\;\nonumber\\
    D(s)&=&\frac{2}{3}\lambda(s,m_K^2,m_\pi^2)\left(1+\frac{2s}{M_\tau^2}\right)|F_+(s)|^2+2\Delta_{K\pi}^2\Bigg|1+\frac{\hat{\epsilon}_S s}{M_\tau(m_s-m_u)}\Bigg|^2|F_0(s)|^2\nonumber\\
    &+&\frac{8}{3}\lambda(s,m_K^2,m_\pi^2)\left[s|\hat{\epsilon}_T|^2\left(2+\frac{s}{M_\tau^2}\right)|F_T(s)|^2-\frac{3s}{M_\tau}\Re e[\hat{\epsilon}_TF_+(s)F_T^*(s)]\right]\,.\label{eq_D(s)}
\end{eqnarray}
\section{Form-factors input}\label{sec:FFinput}
We used the dispersive form factors from Ref. \citen{Gonzalez-Solis:2020jlh}, where one- and two-meson tau decays bind $|\hat{\epsilon}_{S,T}^D|$. We explain our procedure with the simplest $F_+$ case, corresponding to the $\pi\pi$ channel~\cite{Pich:2001pj,GomezDumm:2013sib,Gonzalez-Solis:2019iod}. Other $\pi\pi$ form factors and those for the other channels were worked out in Ref. \citen{Descotes-Genon:2014tla, Miranda:2018cpf,Jamin:2000wn,Jamin:2001zq,Moussallam:2007qc,Boito:2008fq,Boito:2010me,Antonelli:2013usa,Bernard:2013jxa,Rendon:2019awg,Escribano:2013bca,Escribano:2014joa,Gonzalez-Solis:2019lze} and agree well with present data.\\

The di-pion vector form factor is obtained via a dispersion relation with three subtractions:
\begin{equation}
F_+^{\pi\pi}(s)=\mathrm{exp}\left[\alpha_1s+\frac{\alpha_2}{2}s^2+\frac{s^3}{\pi}\int_{4m_\pi^2}^\infty\frac{\mathrm{d}s'}{s'^3}\frac{\delta_+(s')}{s'-s-i0}\right]\,,
\end{equation}
where $\alpha_{1,2}$ are subtraction constants ($F_+(0)=1$ fixes the first one) characterizing slope and curvature of the threshold expansion of $F_+(s)$, where Chiral Perturbation Theory constraints are fundamental~\cite{Weinberg:1978kz, Gasser:1983yg, Gasser:1984gg, Cata:2007ns}. As the phase shift $\delta_+(s)$ is not known everywhere, several approaches are used to estimate the associated uncertainty (see the quoted refs). Around the GeV, the seed for $\tan\delta_+(s)=\frac{\Im m F_+(s)}{\Re e F_+(s)}$ comes from Resonance Chiral Theory ($R\chi T$)~\cite{Ecker:1988te,Ecker:1989yg}, with parameters encoding resonance properties, which are fitted to data. All form factors are required to satisfy the leading QCD asymptotic behaviour~\cite{Lepage:1980fj}.\\

\begin{figure}[h!]
\includegraphics[width=10cm]{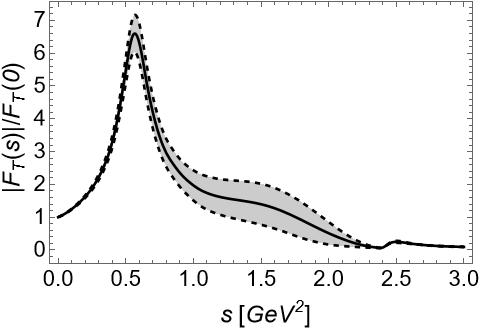}
\centering
\caption{Pion tensor form factor modulus as a function of the $\pi\pi$ invariant mass. The solid curve corresponds to the central value, and the dashed ones covers one standard deviation uncertainties, assuming $|\delta_+(s)-\delta_T(s)|=2\delta_+^{inel}(s)$.}\label{Fig_AbsFT}
\end{figure}

 In the sub-GeV elastic zone, the form factor phase is controlled with minute accuracy~\cite{Garcia-Martin:2011iqs,Caprini:2011ky}. Ref. \citen{Gonzalez-Solis:2019iod}, within the excellent approximation of isospin symmetry, uses these results in the elastic domain, including a number of possible variations in their framework, to gauge reliably systematic uncertainties. The tensor form factor phase, $\delta_T(s)$, thus obtained is compared in fig.~\ref{Fig_deltaFT} to $\delta_+(s)$. Fig.~\ref{Fig_AbsFT} plots $|F_T(s)|/F_T(0)$. Since we ignore the slope parameters, a once-subtracted dispersion relation is used, increasing thus the sensitivity to higher energies. Consequently, the error on $|F_T(s)|$ is larger than for $|F_+(s)|$.\\

\begin{figure}[h!]
\includegraphics[width=10cm]{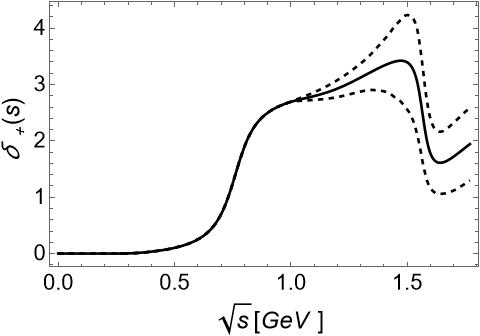}
\centering
\caption{Pion vector form factor phase as a function of the $\pi\pi$ invariant mass. The solid curve represents the central value, and the dashed lines cover the one sigma uncertainties~\cite{Gonzalez-Solis:2019iod}.}\label{Fig_deltaF+}
\end{figure}
\begin{figure}[h!]
\includegraphics[width=10cm]{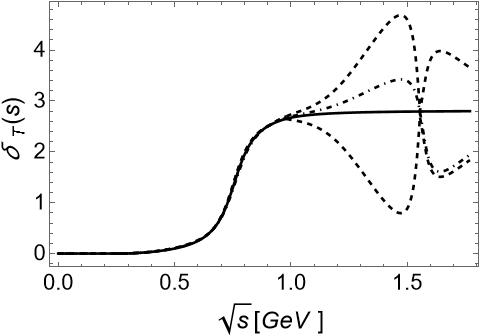}
\centering
\caption{Pion tensor form factor phase as a function of the $\pi\pi$ invariant mass. The solid curve corresponds to the inelastic contributions, and the black dashed ones depict one standard deviation uncertainties, assuming $|\delta_+(s)-\delta_T(s)|=2\delta_+^{inel}(s)$. The central value of $\delta_+(s)$, according to Ref. \citen{Gonzalez-Solis:2019iod} (see fig.~\ref{Fig_deltaF+}), is shown by a black dot-dashed line for reference.}\label{Fig_deltaFT}
\end{figure}
%
%
%
\section{Bounds on the New Physics coefficients}\label{sec:BoundsonEpsilons}
Eq.~(\ref{LEFTLagrangian}) comes from the low-energy limit of the SMEFT Lagrangian. This observation is useful in constraining the imaginary part of the scalar and tensor Wilson coefficients~\cite{Cirigliano:2017tqn, Chen:2021udz}.

For the tensor operators, in the weak basis, one has
 \begin{eqnarray}
&& \mathcal{L}_{SMEFT} \supset [C^{(3)}_{\ell e q u}]_{klmn}(\bar{\ell}^i_{Lk}\sigma_{\mu\nu}e_{Rl})\epsilon^{ij}(\bar{q}^j_{Lm}\sigma^{\mu\nu}u_{Rn})+\mathrm{h.c.}\nonumber\\
&& = [C^{(3)}_{\ell e q u}]_{klmn}[(\bar{\nu}_{Lk}\sigma_{\mu\nu}e_{Rl})(\bar{d}_{Lm}\sigma^{\mu\nu}u_{Rn})-(\bar{e}_{Lk}\sigma_{\mu\nu}e_{Rl})(\bar{u}_{Lm}\sigma^{\mu\nu}u_{Rn})]+\mathrm{h.c.},\;\;\;\;\;\;\;\;\;\;
 \end{eqnarray}
 where the left-handed lepton and quark $SU(2)_L$ doublets are $\ell_L=(\nu_L,e_L)^T$ and $q_L=(u_L,d_L)^T$, while the corresponding singlets are $e_R$ and $u_R$. $SU(2)_L$(family) indices are denoted by latin letters. In the \textit{down} mass basis, one finds
 \begin{eqnarray}\label{eq_SMEFT_MB}
 \mathcal{L}_{SMEFT}\supset [C^{(3)}_{\ell e q u}]_{klmn}[(\bar{\nu}_{Lk}\sigma_{\mu\nu}e_{Rl})(\bar{d}_{Lm}\sigma^{\mu\nu}u_{Rn})-V_{am}(\bar{e}_{Lk}\sigma_{\mu\nu}e_{Rl})(\bar{u}_{La}\sigma^{\mu\nu}u_{Rn})]+\mathrm{h.c.}
 \end{eqnarray}
 In eq.~(\ref{eq_SMEFT_MB}) we are interested in the tau flavor case, $k=l=3$. In the quark sector, $n=1$, and $m=1,2$ will be relevant for the first term ($D=d,s$). In the second one, also $n=2$ will be important (for $D^0-\bar{D}^0$ mixing), so $a,m=1,2$.
 The LEFT and SMEFT coefficients are related at leading order by
 \begin{equation}
 [C^{3}_{\ell equ}]_{33mn}=-2\sqrt{2}G_FV_{uD}\hat{\epsilon}^{D*}_T\,.
 \end{equation}
 The operators in the second term of eq.~(\ref{eq_SMEFT_MB}) contribute to the neutron electric dipole moment ($u$-quark part) and to neutral $D$ meson mixing, constraining strongly $\Im m[\epsilon_T^D]$. As in ref.~Ref. \citen{Chen:2021udz}, two cases will be considered:
 \begin{itemize}
\item If the neutron EDM~\footnote{It is worth remembering that~\cite{Chen:2021udz} $d_n=g_T^u(\mu)d_u(\mu)$, where $g_T^u(2\mathrm{\,GeV})=-0.204(14)$ \cite{Bhattacharya:2015esa, Gupta:2018lvp}, $|d_n|<1.8\times10^{-26}e$ cm at $90\%$ confidence level, C. L. \cite{Abel:2020pzs}.} is mainly contributed by a single $\hat{\epsilon}_T^D$, then
\begin{equation}
d_u(\mu)=-2\sqrt{2}G_F\frac{eM_\tau}{\pi^2}V_{uD}^2\Im m[\hat{\epsilon}^D_T(\mu)]\log\frac{\Lambda}{\mu}\,,
\end{equation}
yielding $|\Im m[\hat{\epsilon}_T^s]|\lesssim4\times10^{-6}$ and $|\Im m[\hat{\epsilon}_T^d]|\lesssim8\times10^{-5}$ for $\Lambda\gtrsim 100$ GeV and $\mu=2$ GeV. The restriction on $D^0-\bar{D}^0$ mixing will then come from $\phi=\Im m\Big [V_{uD}V_{cD}\epsilon_{T(S)}^D\Big]$.
\item If both terms contribute, then the corresponding constriction is
\begin{equation}
d_u(\mu)=-2\sqrt{2}G_F\frac{eM_\tau}{\pi^2}(V_{ud}^2 \Im m[\hat{\epsilon}_T^d(\mu)]+V_{us}^2 \Im m[\hat{\epsilon}_T^s(\mu)])\log\frac{\Lambda}{\mu}\,.
\end{equation}
In this case, the constraint from meson mixing will apply to $\phi=\Im m\Big [V_{ud}V_{cd}\epsilon_{T(S)}^d+V_{us}V_{cs}\epsilon_{T(S)}^s\Big]$.
 \end{itemize}
 We will consider $\phi=\pm\frac{\pi}{4}$~\cite{Cirigliano:2017tqn, Chen:2021udz}.

 For the scalar operators in the SMEFT we have
 \begin{equation}
\mathcal{L}_{SMEFT}\supset [C^{(1)}_{\ell e q u}]_{klmn}\left(\bar{\ell}^i_{Lk}e_{Rl}\right)\epsilon^{ij}\left(\bar{q}^j_{Lm}u_{Rn}\right)+[C_{\ell e d q}]_{klmn}\left(\bar{\ell}^i_{Lk}e_{Rl}\right)\left(\bar{d}_{Rm}q^i_{Ln}\right)+\mathrm{h.c.}\,,
 \end{equation}
 that, in the mass basis, is
  \begin{eqnarray}
\mathcal{L}_{SMEFT}&\supset& [C^{(1)}_{\ell e q u}]_{klmn}\left[\left(\bar{\nu}_{Lk}e_{Rl}\right)\left(\bar{d}_{Lm}u_{Rn}\right)-V_{am}(\bar{e}_{Lk}e_{Rl})(\bar{u}_{La}u_{Rn})\right]\nonumber\\
&+&[C_{\ell e d q}]_{klmn}\left[V_{an}^*\left(\bar{\nu}_{Lk}e_{Rl}\right)\left(\bar{d}_{Rm}u_{La}\right)+\left(\bar{e}_{Lk}e_{Rl}\right)\left(\bar{d}_{Rm}d_{Ln}\right)\right]+\mathrm{h.c.}\;\;\;\;\;\;\;\;\;\;\;
 \end{eqnarray}
 We are again interested in the $\tau$, $k=l=3$ case, and $m,n,a=1,2$. LEFT and SMEFT coefficients are related by ($V_{ub}^*$ part is negligible)
 \begin{equation}
-2\sqrt{2}G_FV_{uD}^* \hat{\epsilon}_S^{D*}=[C^{(1)}_{\ell e q u}]_{33m1}+V_{ud}^*[C_{\ell e d q}]_{33m1}+V_{us}^*[C_{\ell e d q}]_{33m2}
.
 \end{equation}
 The scalar operators do not enter directly the neutron EDM, but they are still indirectly constrained, since the bound on the tensor coefficients affects the scalar ones via renormalization group evolution~\cite{Chen:2021udz}. Keeping for simplicity only $C^{(1)}_{\ell e q u}$~\cite{Chen:2021udz}, $|\Im m[\hat{\epsilon}_T^s]|\lesssim4\times10^{-6}\Rightarrow|\Im m[\hat{\epsilon}_S^s]|\lesssim2\times10^{-3}$ and $|\Im m[\hat{\epsilon}_T^d]|\lesssim8\times10^{-5}\Rightarrow |\Im m[\hat{\epsilon}_S^d]|\lesssim4\times10^{-2}$.
 In the scalar case, the strongest constraints come from $D$ meson mixing, implying~\cite{Chen:2021udz} $\Im m\left[\hat{\epsilon}^D_S\right]\in[-3.1,1.6]\times10^{-4}$ at $95\%$ C.L., that is the dominant one.

 The real parts of the Wilson coefficients are most effectively constrained by analyzing CP-conserving inclusive and exclusive semi-leptonic tau decays~\cite{Chen:2021udz,Garces:2017jpz,Cirigliano:2017tqn,Miranda:2018cpf,Rendon:2019awg,Gonzalez-Solis:2019lze,Gonzalez-Solis:2020jlh,Cirigliano:2018dyk,Cirigliano:2021yto,Arteaga:2022xxy,Escribano:2023seb}.
 We use the results from Ref. \citen{Escribano:2023seb}, which include improved radiative corrections (see also .~Ref. \citen{Arroyo-Urena:2021nil,Arroyo-Urena:2021dfe}), that are
 \begin{equation}
 \Re e[\epsilon_S^d]=\left(1.0^{+0.6}_{-3.4}\right)\times10^{-1}\,,\quad \Re e[\epsilon_T^d]=\left(0.4^{+4.3}_{-4.6}\right)\times10^{-2}\,,
 \end{equation}
 with a correlation coefficient of $0.452$, and
 \begin{equation}
 \Re e[\epsilon_S^s]=\left(0.8\pm0.9\right)\times10^{-2}\,,\quad \Re e[\epsilon_T^s]=\left(0.5\pm0.8\right)\times10^{-2}\,,
 \end{equation}
 almost uncorrelated.
\section{Phenomenological analysis}\label{sec_Pheno}
 We will summarize the numerical analysis~\cite{Aguilar:2024ybr} of the most interesting observables sensitive to CPV using the ingredients introduced in previous sections. The $K_S\pi$ channel has been extensively  discussed elsewhere~\cite{Chen:2021udz,Cirigliano:2017tqn,Rendon:2019awg}. We will present the rate CP and the forward-backward asymmetries for the $\pi^\pm\pi^0$, $K^\pm K_S$ and $K^\pm\pi^0$ channels. Our results for $A_{CP}^{rate}$ will be quoted for the maximum possible signal, corresponding to 
 $\Im m[\hat{\epsilon}_T^{d}] =\pm8\times10^{-5}$ and $\Im m[\hat{\epsilon}_T^{s}] =\pm4\times10^{-6}$ 
 at $90\%$ C.L. For $A_{FB}$, the optimal values of the $\hat{\epsilon}_{S,T}^D$ will be determined by maximizing the figure of merit
 \begin{equation}\label{eq_FOM}
    F(\hat{\epsilon}_{D}^{D},\hat{\epsilon}_{T}^{D}) = \int_{(m_{1} + m_{2})^2}^{M_{\tau}^2}  \Big|A_
    {FB}(s;\hat{\epsilon}_{S}^{D},\hat{\epsilon}_{T}^{D})\Big| \quad  ds\,.
\end{equation}\\
\section{CP violating asymmetries in the \texorpdfstring{$\pi^\pm\pi^0 $} 
channel}
 In this case, the SM contribution is negligible, so using eq.~(\ref{eq_ACPrateNP}) for the NP part, we find
\begin{equation} \label{ACPratepipi}
 A_{CP}^{rate}|_{\pi\pi}\leq3\times10^{-5}\,,
 \end{equation}
which is below the expected sensitivity of current and near-future experiments.\\
 
 The maximal $A_{FB}|_{\pi\pi}$, 
 obtained with the values $ \Re e[\epsilon_S^d]=-3.1\times10^{-2}  $ , $ \Im m[\epsilon_S^d]=-2.7\times 10^{-4} $, $ \Re e[\epsilon_T^d]=-7.9\times 10^{-2}  $, and $ \Im m[\epsilon_T^d]=8 \times 10^{-5}$, is always smaller than $0.1$ in magnitude, and quite close to the SM prediction. It will thus be very difficult to measure, and even more to disentangle a new physics contribution.



%
%
 \subsection{CP violating asymmetries in the \texorpdfstring{$K^\pm K_S$} channel}\label{sec_KK}
This is the most interesting channel (besides $K_S\pi$). Using eq.~(\ref{eq_FullACPrate}) we find
 \begin{equation} \label{ACPrateKK}
 A_{CP}^{rate}|_{KK}=3.8\times10^{-3},
 \end{equation}
 which is dominated by the SM contribution 
 and
 \begin{equation} \label{ACPrateKKNP}
 A_{CP}^{rate}|_{KK,NP}\leq 2.3\times10^{-4},
 \end{equation}
 obtained with the analog equation to (\ref{eq_ACPrateNP}).
Remarkably, a $5\%$ precision in the  $A_{CP}^{rate}$ measurement would already be sensitive to the maximum allowed NP contribution in this channel. This contrasts sharply with the $K_S \pi^\pm$ modes, where the NP contributions is orders of magnitude smaller than the SM one, complicating the explanation of the BaBar anomaly. There are two reasons why this NP contribution is so much larger than for the $K_S \pi^\pm$ channels. First, the bound on the Wilson coefficient is a factor twenty larger here ($|\Im m[\hat{\epsilon}_T^d]|\leq 8\times10^{-5}$ vs. $|\Im m[\hat{\epsilon}_T^s]|\leq 4\times10^{-6}$)
. 
Additionally, the $K_S K^\pm$ channel is fully inelastic, while the main contribution to the $K_S \pi^\pm$ decay width comes from the elastic region, where $A_{CP}^{rate}$ is identically zero. Together, they explain why we get such a large NP effect, that can be up to $\mathcal{O}(5\%)$ in $A_{CP}^{rate}|_{KK}$. This is encouraging for the experimental collaborations, that will target this precision in their measurements. Noteworthy, within $10\%$ accuracy, $A_{CP}^{rate}$ in the $K_S K$ channels should coincide with the $K_S \pi$ result, being a purely experimental test of the BaBar anomaly, that we strongly suggest BaBar and Belle(-II) to do. In the BaBar measurement of this decay \cite{BaBar:2018qry}, the uncertainty of the best measured points is $\sim 3\%$, so these numbers are not unreachable.\\
 
\begin{figure}[h!] 
\includegraphics[width=10cm]{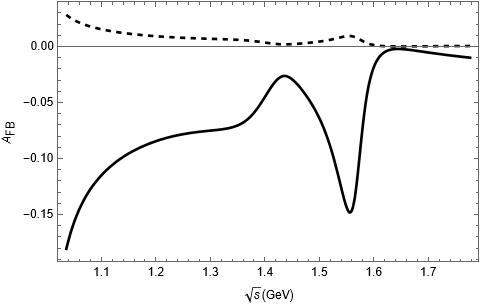}
\centering
\caption{Maximal $A^{\tau\rightarrow K_SK\nu_{\tau}}_{FB}(s) $ (black solid line), corresponding to the Wilson coefficients values $\Re e[\epsilon_S^d]=-3.1\times10^{-2}  $ , $ \Im m[\epsilon_S^d]=-2.7\times 10^{-4} $,   $ \Re e[\epsilon_T^d]=-7.9\times 10^{-2}  $ and  $ \Im m[\epsilon_T^d]=8 \times 10^{-5}$, compared to the SM case (black dashed line).}\label{fig_AFBKK}
\end{figure}

 The maximal $A_{FB}|_{KK}$ is shown by a black solid line in figure \ref{fig_AFBKK}, which is obtained with the same previous set of Wilson coefficients, $\Re e[\epsilon_S^d]=-3.1\times10^{-2}  $ , $ \Im m[\epsilon_S^d]=-2.7\times 10^{-4} $,   $ \Re e[\epsilon_T^d]=-7.9\times 10^{-2}  $ and  $ \Im m[\epsilon_T^d]=8 \times 10^{-5}$. The (unmeasurably small) SM contribution is represented by a black dashed line. The obtained asymmetries, of the order of $-0.1$ in a big portion of the spectrum, should be observable at current and forthcoming facilities.\\

\begin{figure}[h!]

\includegraphics[width=1\linewidth, height=8cm]{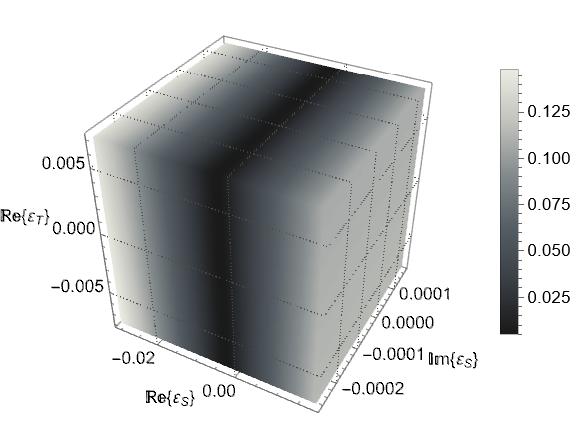}
 
\caption{3D density plot for the FOM described in eq. (\ref{eq_FOM}) on the parameter space, fixing $|\Im m[\hat{\epsilon}_T^d]| = 8\times 10^{-5} $,  for the $KK_S$ decay channel. }\label{3DKK}
\end{figure}

 \subsection{CP violating asymmetries in the \texorpdfstring{$K^\pm\pi^0$} channel}\label{sec_Kpi}
 From eq.~(\ref{eq_ACPrateNP}) we find
 \begin{equation} \label{ACPrateKpi}
 A_{CP}^{rate}|_{K\pi}\leq6\times10^{-7}\,,
 \end{equation}
 again too small to be probed soon.\\
 
 The maximal $A_{FB}|_{K\pi}$ 
 is obtained for the following values of the Wilson coefficients:  $ \Re e[\epsilon_S^s]=2.3\times 10^{-2}  $ , $ \Im m[\epsilon_S^s]=-2.7\times 10^{-4} $,   $ \Re e[\epsilon_T^s]=1.8    \times 10^{-2}  $ and  $ \Im m[\epsilon_T^s]=4 \times 10^{-6}  $. 
 The small difference between this maximal $A_{FB}$ and the corresponding SM contribution seems to prevent any hint of NP soon in this observable
 .\\

 
Varying the Wilson coefficients within their allowed ranges, we find that the maximum FOM value is $0.64$, which is obtained for $\Re e[\epsilon_S^s]=2.3\times10^{-2}  $, $ \Im m[\epsilon_S^d]=-2.7\times 10^{-4} $, $ \Re e[\epsilon_T^d]=-8\times 10^{-3}  $ and  $ \Im m[\epsilon_T^d]=-4 \times 10^{-6}$ (versus $0.45$ in the SM, for vanishing $\epsilon$'s). For this maximal CP violation allowed, measurement (and pinpointing a deviation from the SM) should already be possible.

%
%

For the $K\eta^{(\prime)}$ channels, we obtain non-negligible $A_{CP}^{rate}$ ($\lesssim 10^{-5}$) and large deviations due to NP in $A_{FB}$ and the FOM $F$~\cite{DanielThesis}. Given the fact that these are inelastic channels for $K\pi^0$, it would be very challenging to control the SM prediction at the required level to claim NP using the measured asymmetries for these decay modes.

\section{Conclusions}
There is a very active experimental program for testing the surprising BaBar result for the rate CPV asymmetry in $K_S\pi$ $\tau$ decays. Among these efforts, we highlight the importance of the $\tau\to K_S K \nu_\tau$ channels, that are intimately related to the $K_S\pi$ modes in the dominant source of CPV, coming from neutral Kaon mixing within the SM. Besides, since the $K_S K$ channels lie fully in the inelastic region (where the vector and tensor phases can differ) and the corresponding Wilson coefficients have milder bounds than in $K_S \pi$, the NP effects can reach $\mathcal{O}(5\%)$ in the $KK$ modes, allowing for an allowed clean signal that BaBar, Belle(-II) and the super-tau-charm facility will seek eagerly. 

\section*{Acknowledgements}
D. A. L. A. thanks CONAHCYT/SECIHTI for funding during his Ph. D. The author thanks the organizers for the invitation to give this talk and the pleasant event. Useful comments on the draft of this contribution by my coauthors in Ref. \citen{Aguilar:2024ybr} are also acknowledged.

\section*{ORCID}
\noindent Daniel Arturo López Aguilar- \url{https://orcid.org/0009-0002-5872-0062}

\end{document}